\documentclass{PoS}

\title{Constraining the Sea Quark Distributions Through W$^\pm$ Cross Section Ratio Measurements at STAR}

\ShortTitle{W$^\pm$ Cross Section Ratio Measurements at STAR}

\author{M. Posik\thanks{for the STAR Collaboration}\\
        Temple University, Philadelphia, PA USA\\
        E-mail: \email{posik@temple.edu}}

\abstract{Over the past several years, parton distribution functions (PDFs) have become more precise. However there are still kinematic regions where more data are needed to help constrain global PDF extractions, such as the sea quark distributions $\bar{d}$/$\bar{u}$ near the valence region (Bjorken-x $\approx$ 0.1 - 0.3).~Current measurements appear to suggest different high-x behaviors of these distributions, leading to large uncertainties in global fits. The charged W cross section ratio (W$^+$/W$^-$) is sensitive to the unpolarized $u,\;d,\;\bar{u},$ and $\bar{d}$ quark distributions at large $Q^2$ set by the $W$ mass and could help shed light on this discrepancy. The STAR experiment at RHIC is well equipped to measure the leptonic decays of W bosons, in the mid-rapidity range $\left(|\eta| \leq 1 \right)$, produced in proton+proton collisions at $\sqrt{s}$ = 500/510 GeV. At these kinematics STAR is sensitive to quark distributions near Bjorken-x of 0.16. STAR can also measure the W cross section ratio in a more forward bin ranging from 1.1 $< \eta <$ 2.0, which extends the sea quark sensitivity to higher x. RHIC runs from 2011 through 2013 have collected about 350 pb$^{-1}$ of integrated luminosity, and a 2017 run is expected to provide an additional 400 pb$^{-1}$. Presented here are preliminary results for the 2011-2012 charged W cross section ratios ($\sim$100pb$^{-1}$) and an update on the 2013 charged W cross section analysis ($\sim$250 pb$^{-1}$). }

\FullConference{XXV International Workshop on Deep-Inelastic Scattering and Related Subjects\\
		3-7 April 2017\\
		University of Birmingham, UK}

\begin{document}

\section{Motivation}
Parton distribution functions (PDFs) can be used to describe the internal structure of the proton. The x dependence of these PDFs allows one to distinguish between the intrinsic properties of the proton and what is dominated by QCD radiation. Over the years several global analyses (CT14~\cite{CT14}, MMHT14~\cite{MMHT14}, BS15~\cite{BS15}, etc.) have used the available world data to extract PDF distributions.~One particular measurement that can be used to imporve the PDF extractions is the $\bar{d}/\bar{u}$ ratio.~This ratio has been measured by E866~\cite{E866} and a new preliminary result was recently released by the SeaQuest~\cite{SeaQuest} experiment. These data agree at x < 0.2, but for x > 0.25 the data precision decreases and the trends of the two data sets appear to deviate from each other. More data are needed to help further constrain the sea quark ratio.

While E866 and SeaQuest measure the $\bar{d}/\bar{u}$ ratio through Drell-Yan scattering, W production in $pp$ collisions is also sensitive the sea quarks. The W$^{+}$(W$^{-}$) boson is sensitive to the $\bar{d}$($\bar{u}$) quark, which can be seen in equation~\ref{eq:Wdecay}.  

\begin{equation}
\label{eq:Wdecay}
u  + \bar{d} \rightarrow W^+ \rightarrow e^+ + \nu,\;\; d + \bar{u} \rightarrow W^- \rightarrow e^- + \bar{\nu}. 
\end{equation}

At leading order the charged W cross-section ratio~\cite{Soffer94}, $\sigma_{W+}/\sigma_{W-}$, is directly proportional to the sea quarks as shown in equation~\ref{eq:RW} and probes the sea quark distribution at a larger $Q^2 \sim M^2_W$. 

\begin{equation}
\label{eq:RW}
\frac{\sigma_{W+}}{\sigma_{W-}} \sim \frac{u(x_1)\bar{d}(x_2) + \bar{d}(x_1)u(x_2)}{\bar{u}(x_1)d(x_2) + d(x_1)\bar{u}(x_2)}. 
\end{equation}

\section{Experiment}
The STAR experiment at RHIC~\cite{STAR} is an excellent place to measure the charged W cross section ratio, which was first measured in the 2009 run that collected about 13.2 pb$^{-1}$ of data~\cite{STAR2012}. The charged W cross section ratio was measured in $pp$ collisions at center of mass energy $\sqrt{s} = 500/510$ GeV during the 2011, 2012, 2013, and 2017 running periods. The kinematic reach of STAR allows for a nice complimentary measurement to that of E866 and SeaQuest at a larger $Q^2$ which is set by the $W$ boson mass. In the mid-rapidity region ($|\eta|\le 1$) STAR probes the x range of approximately 0.1 to 0.3. There are several subdetectors used to select electrons/positrons which likely decayed from W bosons, as well as separate their charge: the time projection chamber (TPC)~\cite{TPC}, used for particle tracking, the barrel electromagnetic calorimeter (BEMC)~\cite{BEMC}, used to measure particle energy and for triggering, and the endcap electromagnetic calorimeter (EEMC)~\cite{EEMC}, which is used to estimate background contributions. However ongoing analysis is using the EEMC to extend the charged W cross section ratio measurement in the forward direction, which will result in broadening the x reach of STAR to roughly 0.06$\le$ x $\le$ 0.4. The data sample used in the 2011-2012 analysis totals about 100 pb$^{-1}$ of data, while the 2013 analysis used about 250 pb$^{-1}$ of data. Furthermore the recently completed 2017 running has collected about 350 pb$^{-1}$ of integrated luminosity.      

\section{Results}
The charged $W$ cross section ratio can be measured experimentally as

\begin{equation}\label{eq:Exp-RW}
  \frac{\sigma_{W^+}}{\sigma_{W^-}} = \frac{\left(N^+_O - N^+_B\right)}{\left(N^-_O - N^-_B\right)}\frac{\epsilon^-}{\epsilon+},
\end{equation}

\noindent where $\pm$ corresponds to positively or negatively charged lepton, $N_O$ is the number of events that pass the lepton selection cuts, $N_B$ is the number of background events estimated to be contaminating the data set, and $\epsilon$ is the efficiency at which $W$ events are detected. 

\begin{figure}[!h]
\centering
\includegraphics[width=1\columnwidth]{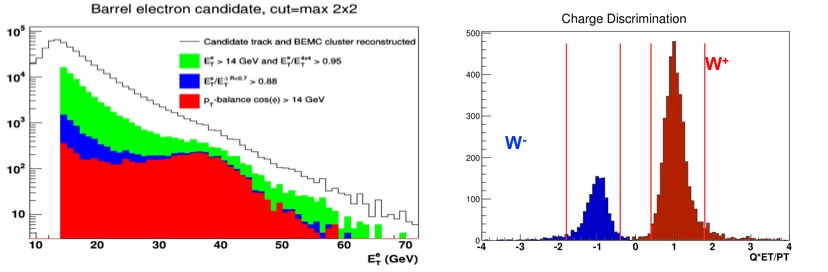}
\caption{Several cuts applied to data that were used to select leptons which likely originated from $W$ decays. Left plot: Application of several isolation cuts including a minimum $E_T$ cut, electron energy ratio cuts, and a signed $p_T$ cut.  Right plot: Projection of the charge separation vs. $E_T$ projected onto the charge separation axis.}
\label{fig:cuts}
\end{figure}

The 2011 and 2012 data sets have been analyzed and preliminary results have been released~\cite{Posik15}. The electrons and positrons from W leptonic decay candidates are selected using methodologies previously developed by STAR~\cite{STAR2012}. The selection process includes placing a cut (25 GeV $\le E_{T} \le$ 50 GeV) which selects candidates populating the Jacobian peak near $E_T \sim M_W$/2, matching high $p_T$ tracks to BEMC clusters, applying several electron/positron isolation cuts based on energy that is deposited in the BEMC, and taking advantage of the large missing momentum due to the decay neutrino going undetected. The left plot of Fig.~\ref{fig:cuts} shows the successive application of various cuts, with the black histogram representing the electron/positron candidates. The green histogram applies a cut requiring $E_T >$ 14 GeV and that 96\% of the electron/positron $E_T$ is contained with in a 4x4 BEMC cluster. The dark blue histogram then applies a cut that requires the 88\% of the electron/positron $E_T$ to be contained in a cone with a radius $\sqrt{\phi^2 + \eta^2} = 0.7$, where $\phi$ is the azimuthal angle and $\eta$ is the electron pseudo-rapidity. The signed-$p_T$ cut, which quantifies the amount of missing transverse momentum, is applied and shown by the red histogram. As each isolation cut is applied a decrease in the background, which populates the kinematic region $E_T < 25$ GeV, and an enhancement in the peak near the Jacobian peak are observed. A charge separation cut is then applied to the candidates to select electrons and positrons that were likely produced from W$^-$ or W$^+$ decays, shown in the right plot of Fig.~\ref{fig:cuts}. The charge selection cuts are indicated by the red lines.

\begin{figure}[!h]
\centering
\includegraphics[width=1\columnwidth]{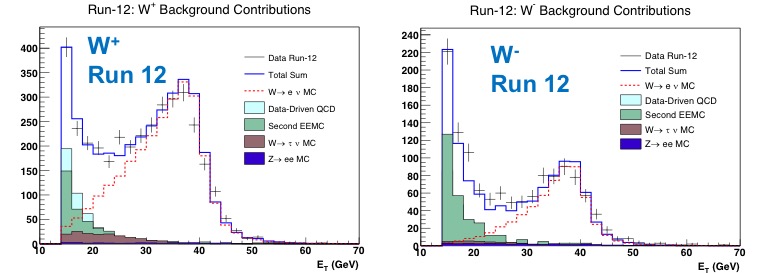}
\caption{Background contributions for W$^{+}$ (left) and W$^{-}$ (right) distributions for the 2012 data set.}
\label{fig:Back}
\end{figure}

Figure~\ref{fig:Back} shows the W$^+$ and W$^-$ background contributions for the 2012 data set. The background contributions include events from $W\rightarrow \tau + \nu$, $Z\rightarrow ee$, QCD, and second EEMC. The QCD and second EEMC backgrounds were estimated using the data, while the other background contributions are computed from Monte Carlo. An estimate of the amount of QCD background that is present in the data is determined from the $E_T$ distribution that fails the signed-$p_T$ cut. This distribution is dominated by QCD type events. The second EEMC background is an estimate of the background caused by an escaping jet's $p_T$ being misidentified as the neutrino's missing $p_T$. Also included in the figure is the Monte Carlo simulation of the W decay signal (based on Pythia 6.4.22~\cite{Pythia} and GEANT~\cite{GEANT}), and comparison of Monte Carlo signal with background contributions added to the 2012 data. When the final analysis cut requiring $E_T > 25$ GeV is applied, there is little background contamination remaining. The 2011 data set showed a similar behavior to the 2012 data set.

\begin{figure}[!h]
\centering
\includegraphics[width=1\columnwidth]{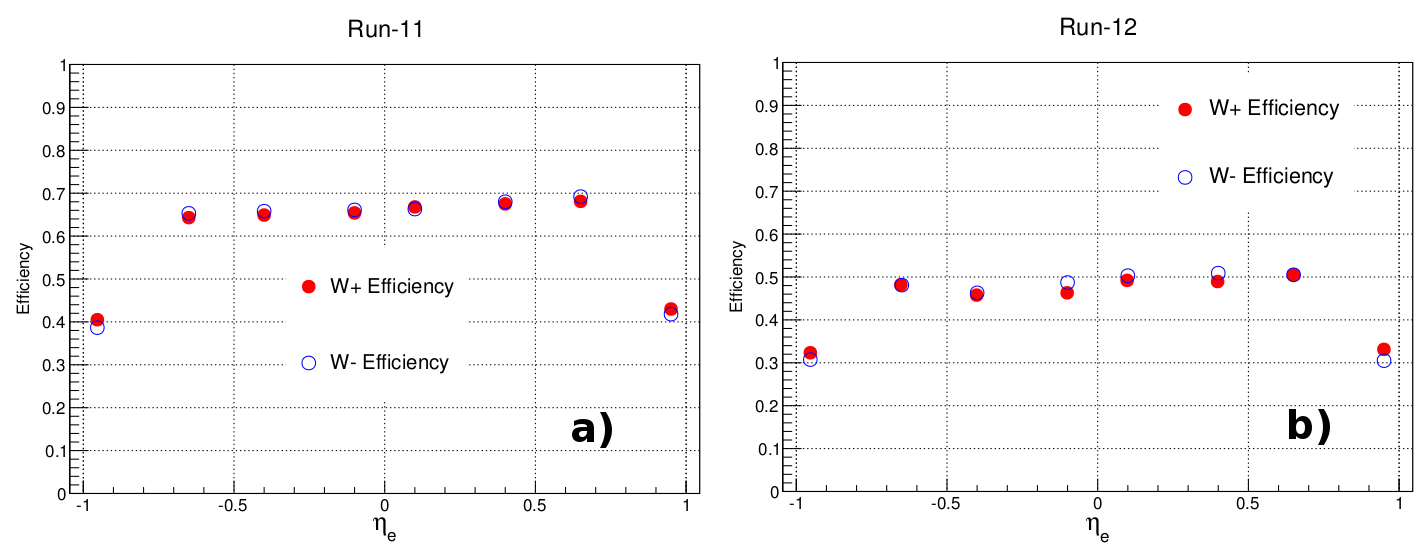}
\caption{W$^{\pm}$ detection efficiency measured in 2011 (left plot) and 2012 (right plot) data sets. }
\label{fig:Eff}
\end{figure}

Monte Carlo simulations were used to determine the W$^\pm$ detection efficiencies. The 2011 and 2012 W efficiencies are shown in Fig.~\ref{fig:Eff} and include all cut and acceptance efficiencies. In the 2012 running RHIC increased the average instantaneous luminosity available per beam fill, which resulted in lower W detection efficiencies relative to the 2011 efficiencies due to more pile-up events in the TPC. The W detection efficiency showed a very small dependence ($\sim$ 1-2\%) on the charge of the W, thus the $\frac{\epsilon^-}{\epsilon+}$ factor in equation~\ref{eq:Exp-RW} will have a negligible effect on the charged W cross section ratio. 

\begin{figure}[!h]
\centering
\includegraphics[width=1\columnwidth]{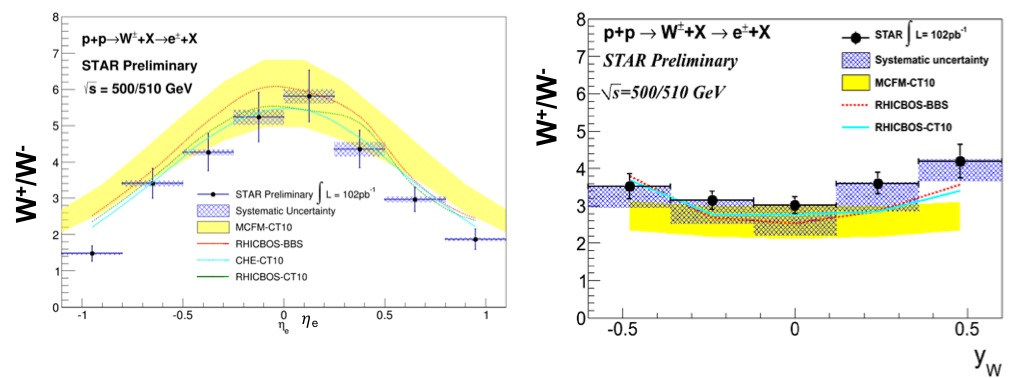}
\caption{Preliminary charged W cross section ratio as a function of electron pseudo-rapidity (left) and W rapidity (right) plots for the 2011-2012 data sets. }
\label{fig:RW}
\end{figure}

Using the lepton W decay candidates and equation~\ref{eq:Exp-RW}, the charged W cross section ratio can be computed. Figure~\ref{fig:RW} (left plot) shows the preliminary charged W cross section ratio plotted as a function of lepton pseudo-rapidity measured from the 2011-2012 data sets, first released at DIS 2015~\cite{Sal15,Posik15}. The error bar on the data points in Fig.~\ref{fig:RW} represent the statistical uncertainty, while the shaded boxes correspond to the systematic uncertainty. The yellow band and colored curves serve as a comparison to different PDF sets~\cite{CT10nlo,BBS} and theory frame works~\cite{MCFM,RHICBOS01}. These data have well controlled systematic uncertainties and are dominated by statistical uncertainties. Through the use of Monte Carlo, the W boson kinematics can be reconstructed using a technique that was developed at FNAL~\cite{Wboson-Fermi} and LHC~\cite{Wboson-LHC}. The neutrino transverse momentum is reconstructed from the missing $p_T$ via Monte Carlo, and then the neutrino's longitudinal momentum can be reconstructed from the decay kinematics~\cite{Sal14,Sal15}. The preliminary 2011-2012 charged W cross section ratio is shown as a function of the W rapidity in the right plot of Fig.~\ref{fig:RW}. One benefit of using the reconstructed W kinematics is that you remove the angular dependence of the decayed electron/positron. However, using the fully reconstructed W kinematics does lead to larger systematic uncertainties, primarily associated with smearing effects introduced during the reconstruction process, as compared to using the electron pseudo-rapidity~\cite{Sal14,Sal15}. 

\section{Analysis Update}
\begin{figure}[!h]
\centering
\includegraphics[width=1\columnwidth]{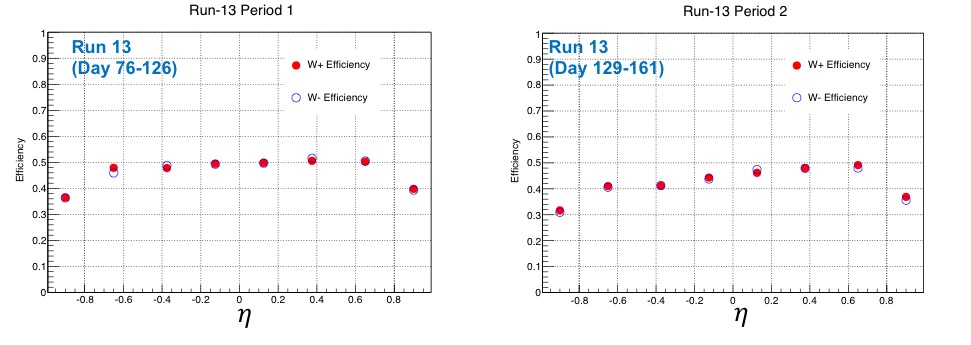}
\caption{W$^\pm$ efficiencies measured with the 2013 data set before (left plot) and after (right plot) the insertion of a new subdetector.}
\label{fig:13Eff}
\end{figure}

The 2013 data set analysis is ongoing. A significant update implemented in the 2013 data analysis, which ran in a higher pile up environment than the 2012 run, was switching to an updated tracking algorithm used to find tracks in the TPC. The updated tracking algorithm implemented a new track-seed finder which looks throughout the entire TPC sector for seeds. The previous tracking algorithm looked for seeds starting at the outer part of the TPC sector and extended tracks to the inner part of the TPC. The updated tracking algorithm led to an increased W detection efficiency in a high pile-up environment. As a result the 2013 W detection efficiencies were found to be in agreement with the 2012 efficiencies, and can be seen in Fig.~\ref{fig:13Eff}. The left plot of Fig.~\ref{fig:13Eff} shows the W$^\pm$ efficiencies before the insertion of the heavy flavor tracker (HFT, $\sim$ 3\% radiation length)~\cite{HFT}, an additional subdetector, into the TPC sensitive volume and the right plot shows the efficiencies after the HFT insertion into STAR, which caused a slight dip in the efficiency in the region $\eta < 1$.

The BEMC was also calibrated using the 2013 $pp$ data at $\sqrt{s} = 510$ GeV and was applied to the 2013 analysis. There have already been two preliminary W single-spin asymmetry results released that incorporate these updates and calibrations~\cite{WAL2016,WAL2017}.  

Analysis work is also being done on the EEMC to obtain a charged W cross section ratio measurement from that detector. A measurement using the EEMC would extend the x reach of the STAR measurements to an x range of roughly 0.06 to 0.4, as the EEMC covers the region $1.1\le \eta \le 2$.

\section{Summary}
STAR has measured the charged W cross section ratio using W production in $pp$ collisions at $\sqrt{s} =$ 500 GeV and 510 GeV. These measurements will provide additional high $Q^2$ data that is sensitive to the sea quark ratio in the kinematic range of about 0.06 $\le$ x $\le$ 0.4, which should help constrain the sea quark distribution. Furthermore, the STAR results will serve as a complementary measurement to E866 and SeaQuest. Preliminary charged W cross section ratio results for the 2011 and 2012 data sets, which account for about 100 pb$^{-1}$ of data, have already been released both as a function of electron pseudo-rapidity and W rapidity. 

The 2013 data set, which accounts for about 250 pb$^{-1}$ of data is currently under analysis. This data set benefits from an updated tracking algorithm, which allows for more efficient track reconstruction in a higher pile up environment. The 2013 analysis is also working on obtaining a charged W cross section result from the EEMC, which would broaden the x-range that the STAR measurements are sensitive to, since the EEMC sits more forward in $\eta$.

Additionally, the 2017 STAR run, which collided $pp$ at $\sqrt{s} = 510$ GeV, just finished running and collected $\sim$ 350 pb$^{-1}$ of integrated luminosity. These data are still in the very early calibration and production mode, but will ultimately be added to the data collected in 2011,2012, and 2013.

\end{document}